\documentclass[conference]{IEEEtran} 
\usepackage{amsmath}
\usepackage{bm}
\usepackage{eurosym}

\usepackage{url}
\usepackage{comment}
\usepackage{amssymb}
\usepackage{amsfonts}
\usepackage{cite}
\usepackage{graphicx}
\newcommand{\erf}{\text{erf} } 

\begin{document}

\title{Survival Models for the Duration of Bid-Ask Spread Deviations}

\author{\IEEEauthorblockN{Efstathios Panayi}
\IEEEauthorblockA{Department of Computer Science\\
University College London\\
London, WC1E 6BT\\
Email: efstathios.panayi.10@ucl.ac.uk}
\and
\IEEEauthorblockN{Gareth Peters}
\IEEEauthorblockA{Department of Statistics\\
University College London\\
London, WC1E 7HB\\
Email: garethpeters78@gmail.com}}

\maketitle

\begin{abstract}
Many commonly used liquidity measures are based on snapshots of the state of the limit order book (LOB) and can thus only provide information about instantaneous liquidity, and not regarding the local liquidity regime. However, trading in the LOB is characterised by many intra-day liquidity shocks, where the LOB generally recovers after a short period of time. In this paper, we capture this dynamic aspect of liquidity using a survival regression framework, where the variable of interest is the duration of the deviations of the spread from a pre-specified level. We explore a large number of model structures using a branch-and-bound subset selection algorithm and illustrate the explanatory performance of our model. 
\end{abstract}


\section{Introduction}

Market liquidity is considered a desirable characteristic of financial markets, as in liquid markets we generally expect to observe fewer abrupt changes, or `jumps', in terms of either the security price or the volume available for that security. In such markets, participants can both build positions in these securities, and liquidate them, without incurring substantial execution costs. For some investors, liquidity is the most important decision-making criterion in selecting the markets and assets they would like to invest in, and is a central concept that quantifies the quality of particular securities markets. 

We focus on activity in the limit order book (LOB), the central matching mechanism in use in over half of the world's stock exchanges today \cite{rocsu2009dynamic}. It collects all the buying and selling interest in a particular stock and presents an aggregation of these orders to every market participant. \cite{rocsu2009dynamic} suggests that limit orders are submitted at different levels, in order for market participants to protect themselves from adverse selection, that is, execution of their order against a trader with superior information. Thus, understanding the dynamics of the LOB is important and it can help lower trading costs, through the design of a schedule for the execution of large buy or sell orders. 

%
%

A large number of orders resting in the LOB come from market making activity, and it is precisely this activity that we would expect to replenish the LOB after a shock (for example, after a large market order, or a series of cancellations). In this setting, patient traders may want to wait until this replenishment occurs before placing an order. This would reduce the trading costs they would have to incur, by crossing a large bid-ask spread for example. However, a snapshot of the LOB, such as that presented on traders' terminals, is not informative about the time such traders would have to wait.  

In this paper we propose a model for the time required for the spread to recover after a shock. We measure the duration of the deviation of the spread from a pre-specified level and relate it to a number of instantaneous and lagged covariates obtained from the LOB. This is achieved through the development of a survival regression framework. We explore a very large number of model structures, and fit the model to a four month dataset of LOB data from the Chi-X exchange. We can thus identify covariates that have explanatory power over time. 

Our survival regression approach is related to the model of Lo et al. \cite{lo2002econometric}, who had similarly utilised an accelerated failure time formulation. However, their model was for the lifetime of individual limit orders, while ours is for the deviation of the spread and as such is more informative about the liquidity regime. In addition, our regression model incorporates more information about both instantaneous and lagged LOB structural variables, and we show that these have substantial explanatory power in explaining our observations. 

The contribution of this paper is in developing a new approach to modelling the relationship between the intra-day structure of the LOB and the duration of spread deviations from a pre-defined threshold level. In this context, we show that incorporating LOB covariates in a survival regression framework can explain a substantial part of the variation in the spread duration deviations. We also explain how one can use a branch-and-bound algorithm to explore different regression model structures, in order maximise this explanatory power.  

For each feature of the LOB considered as an explanatory variable, we evaluate whether an inter-day stationarity assumption is suitable over our sample period. We find that it is not, and thus fit the model separately to each daily dataset. In addition, we investigate the relative increase in the explanatory power as we increase the number of covariates in the model. Finally, we provide an interpretation for the contribution of the covariates that form part of the optimal models in explaining the variation in the observed durations of spread deviations.



The remainder of this paper is structured as follows: Section \ref{sec:LOB} explains the operation of the LOB and describes features of the dataset used in this paper. Section \ref{sec:relwork} explains how our work relates to the literature in liquidity modelling and survival analysis. Section \ref{sec:model} formally defines the observation random variables, the LOB covariates and the survival model. Section \ref{sec:res} presents our results, in terms of the explanatory power of the model and Section \ref{sec:conc} concludes.


\section{Limit order book data}
\label{sec:LOB}
The central limit order book is a centralised system that displays the trading interest in a particular stock on a given trading venue. Market participants are typically allowed to place two types of orders on the venue: Limit orders, where they specify a price over which they are unwilling to buy (or a price under which they are unwilling to sell), and market orders, which are executed at the best available price. Market orders are executed immediately, provided there are orders of the same size on the opposite side of the book. Limit orders are only executed if there is trading interest in the order book at, or below (above), the specified limit price. If there is no such interest, the order is entered into the limit order book, where orders are displayed by price, then time priority. 

\begin{figure}[!ht]
\begin{center}
\includegraphics[width=0.48\textwidth]{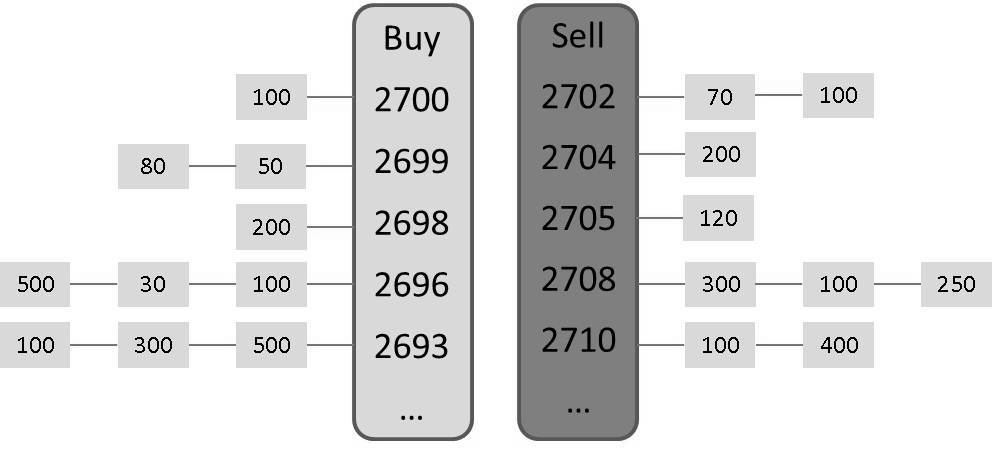}
\vspace{-1em}
\caption{An example of the state of the LOB}
\label{fig:LOB}
\end{center}
\end{figure}

Figure~\ref{fig:LOB} shows an example snapshot of the order book for a particular stock, as traded on the Chi-X exchange, at a particular instance of time. A market order to buy 200 shares would result in 3 trades: 70 shares at 2702, another 100 shares at 2702 and the remaining 30 at 2704. A limit order to sell 300 shares at 2705, on the other hand, would not be executed immediately, as the highest order to buy is only at 2700 cents. It would instead enter the limit order book on the right hand side, second in priority at 2705 cents after the order for 120 shares which is already in the book.  

We study a dataset from the Chi-X exchange (prior to its merger with BATS), for the period between the 2nd of January and the 27th of April 2012. Our 82 day sample contains information relating to the trading of stock Alstom SA, a French energy and transportation company that is part of the CAC40 index. This consists of all timestamped limit order submissions, executions and cancellations within normal business hours. The exchange has both a visible and a hidden order book and orders are routed to each book according to the type and size of the order. The visible book supports the submission of a number of order types \footnote{\url{http://www.chi-xeurope.com/document-library/chi-x-europe-exchange-guide-v2-7-f.pdf}}, however, these are all converted by the exchange matching engine into time stamped limit order submissions, executions and cancellations, and this is what is contained in our dataset.  


\begin{figure}[!ht]
\begin{center}
\includegraphics[width=0.48\textwidth]{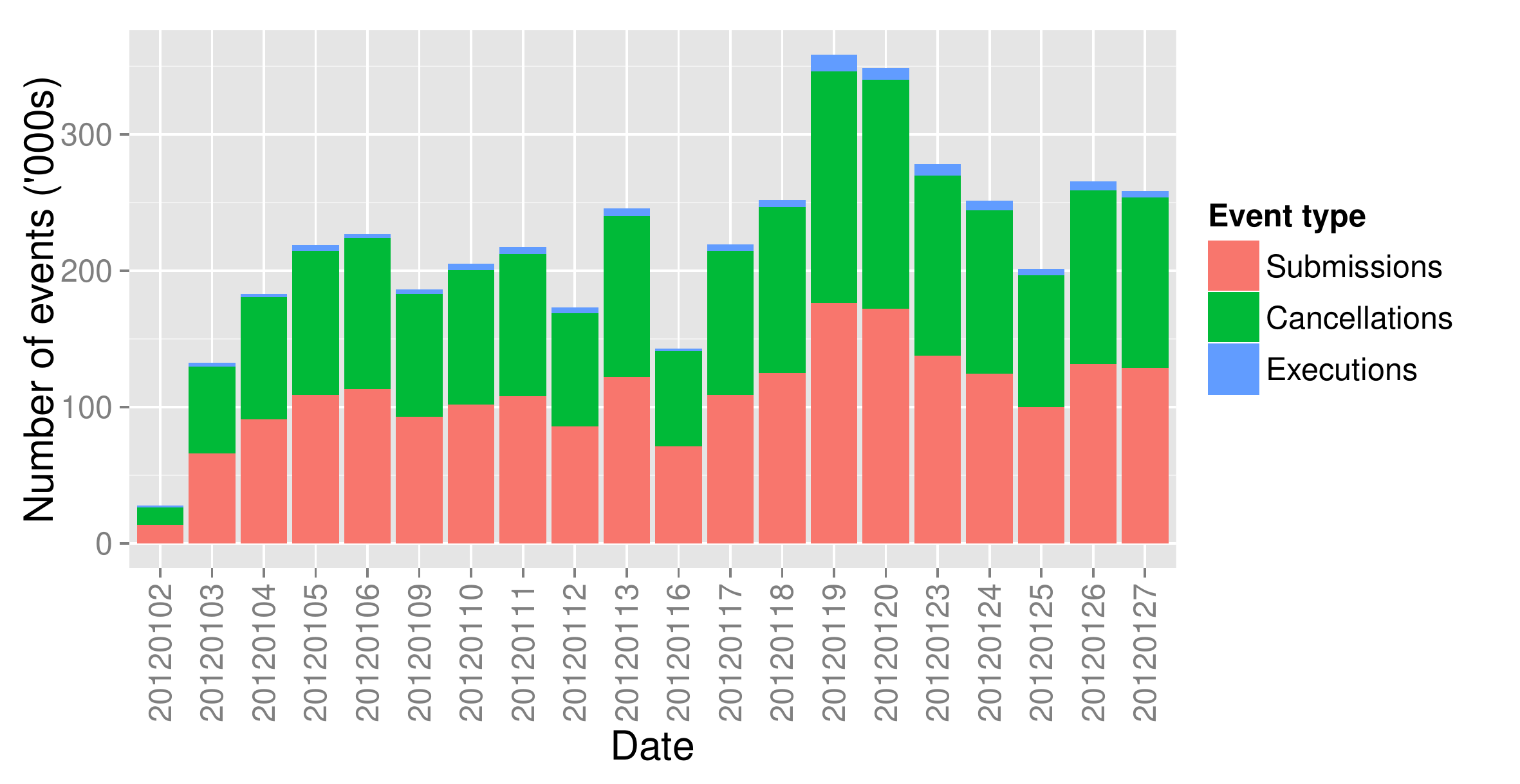}
\vspace{-1em}
\caption{Total trading activity in the LOB for Alstom SA on the Chi-X trading venue over January 2012.}
\label{fig:activity}
\end{center}
\end{figure}

The activity in our dataset is characterised by a relatively low proportion of trades, compared to the number of limit order submissions and cancellations. This is a feature of modern financial markets, and there are a number of factors contributing to this. For example, there are the activities of market makers, who need to price their bids and offers appropriately, so as to avoid the risk of adverse selection \cite{rocsu2009dynamic}. Some proprietary trading algorithms may also produce a large number of cancellations, due in part to `chasing' the current market price for the asset, or seeking `latent' liquidity (i.e. that is available, but not displayed) \cite{hasbrouck2009technology}. There are also orders submitted very far from the top of the order book, and as such, have a very small probability of execution \cite{zovko2002power}. For the dataset considered in this paper, we generally have between 150,000 and 300,000 daily events, of which only between 1.5\% and 5\% are executions, as indicated in Figure \ref{fig:activity}. 

\section{Related work}
\label{sec:relwork}
\subsection{Bid-ask spread literature}



The bid-ask spread, or the difference between the highest bid and lowest offered price in the LOB, represents the cost that an investor must incur in order to be guaranteed immediate execution, i.e. by crossing the spread with a market order.  In the financial literature, early models focusing on quote-driven (dealer) markets had attributed the variation of the spread on inventory holding costs \cite{amihud1980dealership} and the risk of adverse selection \cite{easley1987price}, while Huang and Stoll \cite{huang1997components} also find a large order processing component. Affleck-Graves et al. \cite{affleck1994trading} then found differences in the breakdown of the spread into these components in quote and order-driven markets.


The intra-day variation of the spread in different markets has different properties: For example, while Chan et al. \cite{chan1995market} found a declining intra-day spread for NASDAQ securities, Abhyankar et al. \cite{abhyankar1997bid} found a U-shape pattern. Chordia and Roll \cite{chordia2000commonality}  study of timeseries variations in the bid-ask spread. They found that (long-term) liquidity is influenced by factors such as interest rates, market volatility and seasonal effects.

Our contribution to this literature is in introducing a time component to existing considerations about the variation of the spread. That is, while we do not model the absolute level of the spread, we propose a model with explanatory power about the duration of the deviation of the spread from a specified threshold level. We also focus on much shorter time intervals than those considered thus far, typically in the order of milliseconds. This is necessary in markets which have an increasing representation of high frequency trading firms \cite{hasbrouck2013low}, and where the order flow of such firms has an impact on market quality \cite{brogaard2010high}. Our model is furthermore distinguished from the work attributing the spread to various theoretical determinants, in that it evaluates the explanatory power of covariates without proposing an explicit theoretical stochastic model of the LOB.       

%
\subsection{Survival models }

Survival analysis is a technique used to model the lifetimes of certain variables of interest that are usually subject to competing risks. It has been used in the past to predict financial distress (e.g. \cite{laitinen2005survival,gepp2008role}), to examine the lifetimes of hedge funds \cite{gregoriou2002hedge} and to model the time-to-exit of venture capital firms from their investments \cite{giot2007ipos}. Deville and Riva \cite{deville2007liquidity} used a similar approach, with a Weibull specification, to model the duration of the deviation from put-call parity of the French index options market. They found that the `time-to-no-arbitrage' is systematically linked to liquidity-related variables, such as the traded volume on the options or index markets.

To the best of our knowledge, our paper is the first to examine fluctuations in the spread explicitly using a survival model. However, there has been related work in modelling the lifetime of individual limit orders (that is, the time between execution and cancellation), with early models by Al-Suhaibani and Kryzanowski \cite{al2000exploratory} for the less liquid Saudi stock market, and by Cho and Nelling \cite{cho2000probability} for the New York Stock Exchange. In developing our survival model for the duration of the deviations of the spread, we note that we learn from the survival regression framework proposed in Lo et al. \cite{lo2002econometric} what could be a reasonable subset of covariates (transforms) of the LOB structure to consider in our model. We also employ the accelerated failure time survival formulation used in their model. 

There are two major differences in our approach, compared to that of Lo et al. \cite{lo2002econometric}: Firstly, they handle censoring differently, as they consider cancelled orders to be censored observations, or failed executions. This may be due to the fact that they study a dataset from a brokerage firm that served mainly institutional clients, with a much lower proportion of cancellations. Secondly, they only consider explanatory variables that take into account the top of the order book (highest bid and lowest offer price, as well as the volume at those levels), whereas we take into account the first 5 levels on both the bid and ask side, as well as lagged versions of those variables. We demonstrate the importance of new features not previously utilised in this study. 

A related study by Chakrabarty et al. \cite{chakrabarty2006competing} develops a competing risk framework to further distinguish between the lifetimes of executed or cancelled limit orders in a LOB. In this survival regression study they also utilise the same covariates employed in \cite{lo2002econometric}. In our paper, we do not use a competing risk framework, as for us it not important to make a distinction regarding whether the spread recovered to a former level due to the arrival of limit orders to buy or limit orders to sell.

\section{Spread deviations and model formulation}
\label{sec:model}
\begin{figure}[!ht]
\begin{center}
\includegraphics[width=0.48\textwidth]{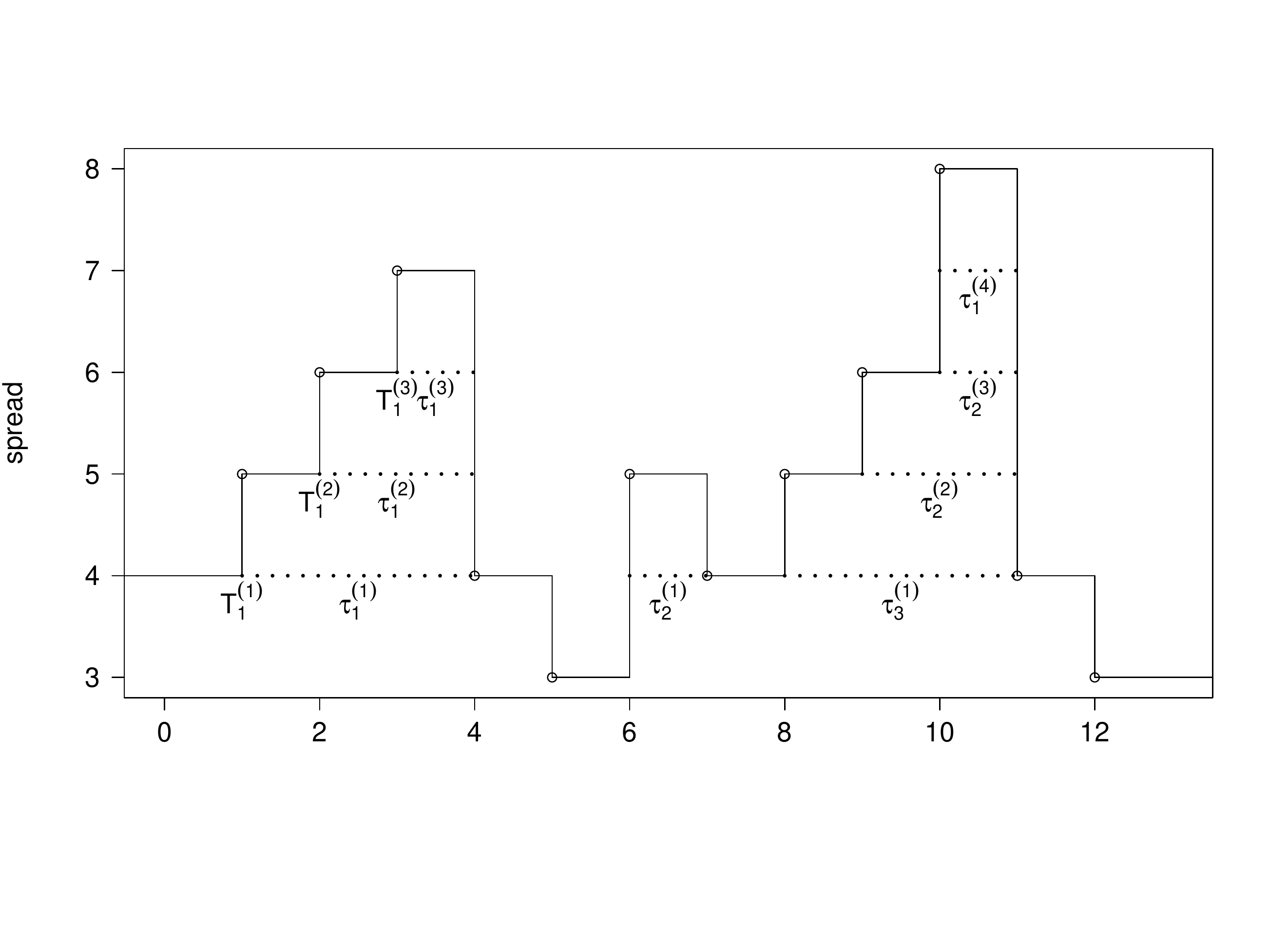}
\vspace{-4em}
\caption{An example of the durations of spread deviations above four different thresholds.}
\label{fig:spreadexceedexamplemultiple}
\end{center}
\end{figure}

Figure \ref{fig:spreadexceedexamplemultiple} shows an example of the durations of deviations above threshold levels of the spread, the quantity we are modelling in our survival regression approach. We show multiple levels here, to note that there is flexibility in the choice of the liquidity threshold, for example as a quantile of some historical distribution of the liquidity measure. The recovery of the spread to (say) its historical median value may be important to a brokerage house or a large fundamentals trader executing a large order, while a regulator may be more interested in the duration of more infrequent events, where the spread reaches very high quantile levels of the empirical distribution. 

We note that in collecting our observations from our dataset, the granularity of the millisecond-stamped transactions and the speed of intra-day LOB revisions are such that we often have deviations of duration 0. As this would cause problems with the estimation of our model, we set the duration of these deviations to 0.1 ms (100 microseconds), which is in the order of the smallest round-trip time for messages to the exchange. 

\subsection{Notation and definitions of LOB variables and survival model components}
\label{definitions}
In general we will reserve upper case letters to denote random variables, bold for random vectors and lower case letters for the realizations of these random variables and vectors. In addition, we utilise the following notation for a single asset, on a single trading day:
\begin{itemize}
\item $a$ denotes the ask, $b$ denotes the bid.

\item $P_{t}^{b,i}\in\mathbb{N}^{+}$ denotes the random variable for the limit price of the $i^{th}$ level bid at time $t$ in tick units. A `level' is defined as one in which there is at least one resting limit order at time $t$.

\item $P_{t}^{a,i}\in\mathbb{N}^{+}$ denotes the random variable for the limit price of the $i^{th}$ level ask at time $t$ in tick units.

\item $\bm{V}_{t}^{b,i}\in\mathbb{N}^{n}$ denotes a column random vector of orders at the $i^{th}$ level bid at time $t$.

\item $S_t=P^{a,1}_{t}-P^{b,1}_{t}$ denotes the random variable for the spread at time $t$.

\item $c$ denotes the threshold level of the spread, defined in the same units as the spread $S_t$. $c$ is deterministic and constant over time.

\item $T_i$ denotes the $i$-th random time instant in a trading day that the spread $S_{t}$ exceeds the threshold $c$. Formally, we define $T_{i} = \inf\left\{t: S_{t} \geq c, \; t \geq T_{i} , \; t>T_0  \right\}$, where $T_0$ denotes the start of the observation window (1 minute after the start of the trading day). 

\item $\tau_i$ will denote the duration of time in ms, relative to the exceedance event $T_i$, that the liquidity measure $S_{t}$ remains above the threshold $c$. These are the response random variables which correspond to the durations of spread deviations. We impose an upper bound $\tau_i \leq T_{D}-T_i$, where $T_D$ denotes the end of the observation window (1 minute before the end of the trading day). If a return of the liquidity measure $S_{t}$ to the threshold $c$ has not occurred by time $T_D$, we consider the observation censored. Censored observations must be accounted for separately in the model estimation, as explained in this section.

\end{itemize}

Using this notation, we define the $P_t^m=\frac{(P_{t}^{a,1}+P_{t}^{b,1})}{2}$ as the random variable of the quoted midpoint, or mid price. In addition, we denote the total volume available at for example the $i$-th bid level by $TV_{t}^{b,i}=\bm{1}_n^T\cdot \bm{V}_{t}^{b,i}$, where $\bm{1}_n$ is a column vector of 1s. We then represent an incoming buy limit order, with information relating to time, price and size and an order id as $\bm{l}^b=(l_t,l_p,l_s,\mathrm{id})$.

For the threshold $c$, in this paper we take all observations of the spread on the 2nd of January 2012. Using these observations, we construct the empirical distribution and define the threshold to be the median of that distribution, which corresponds to \euro 0.03, or 3 cents. 

\subsection{Observations}

Based on these definitions, we can now formally define the observation random variables to illustrate the survival regression framework we develop. In the regression framework we develop we would then select a threshold level $c$ as described above, and define the durations of spread deviations as: 
\begin{equation*}
\label{eq:tedspread}
\tau_i=\inf\left\{t~:~p^{a,1}_{T_i+t}-p^{b,1}_{T_i+t}\leq c, \; t \geq T_{i}, \; t>T_0\right\}
\end{equation*}

We consider here $c$ to be given by the median, and thus  $\tau_i$ is a random variable that represents the duration of the deviation from the `normal' level of liquidity. To understand how such deviation events are generated, the `birth' of the deviation at time $T_i$ would have come either from a market order or from cancellations at the top of the book that had removed one or more levels of the bid or ask. In the analysis we present here, we do not distinguish between the two origins of deviations, but our model is flexible enough to accommodate this easily, e.g. by using a dummy variable to indicate this origin.

The subsequent `death' at time $T_i+\tau_i$ would result from limit orders arriving inside the spread, such that the new spread is at most $c$:
\begin{itemize}
\item $l_p \geq p^{a,1}_{T_i+\tau_i}-c$ for a limit order to buy; or
\item $l_p \leq p^{b,1}_{T_i+\tau_i}+c$ for a limit order to sell
\end{itemize}
where $l_p$ is the price of the incoming limit order.

\subsection{Survival analysis for modelling durations}
\label{subsec:surv}
Survival analysis is a method used to model the time until a particular event occurs, such as the failure of some component or the death of an individual. It is useful in that:

\begin{itemize}
\item It can be used in situations where there can be censored observations, i.e. where the event of interest does not occur during the observation period. For example, when studying the lifetime of mechanical components, we may have that at the end of the observation period, some of the components are still working, in which case the only information we have about them is that their lifetime exceeds the observation period.
\item It can be incorporated into a regression framework, and we can thus explain some of the variation in the variable of interest through explanatory covariates. 
\end{itemize}

In this chapter, we intend to use the technique to model the duration of the deviations of the spread from the (median) threshold value. Let us assume that these observations have an associated probability density function $f(t)$ and cumulative density function $F(t)=P(\tau < t)=\int_{0}^{t}f(t)dt$. From these, we can calculate the survival function $S(t)=1-F(t)$, i.e. the probability that the deviation still holds after a duration $t$, and hazard function $h(t)=\frac{f(t)}{S(t)}$, which is the instantaneous rate of death, given that it had survived up to that point. 

If all observations were uncensored (and independent and identically distributed), then we could simply estimate the model via standard maximum likelihood estimation, where for a given parameter vector $\theta$, the likelihood function is $L(\theta|\tau_1 \ldots \tau_n)=f(\tau_1 \ldots \tau_n|\theta)=\prod_{i=1}^n f(\tau_i|\theta)$. However, because of the presence of censored observations, the calculation of the likelihood function has to be adapted to reflect this.

For a censored observation, we only know that the lifetime $\tau_i$ exceeds the maximum observation time $T_D-T_i$, as we assume that censoring is non-informative (that is, the time of censoring is independent of the time of failure). The contribution to the likelihood of this event is then 

\begin{equation}
L_i = S(T_D-T_i)
\end{equation} 

If we assume for these observations that $T_i$ is independent of $T_d-T_i$, we can then obtain the likelihood function as follows:

\begin{center}
$L=\prod_{i=1}^nL_i=\prod_U f(\tau_i) \prod_C S(T_d-T_i)$
\end{center}
where $U$ and $C$ are uncensored and censored observations, respectively. 

In practice, for a given fixed threshold $c$, once the $i$-th exceedance at time $T_i$ occurs, there is no guarantee that the liquidity process would ever return back through this threshold within the trading day. We do, however, assume that given enough time, the event of interest (i.e. the liquidity measure returning below the threshold) would eventually occur i.e. the liquidity process is mean reverting. Without this assumption, the density we specified for the survival times $f$, that models the distributions of the durations, would be improper, as it would not normalise to unity on its support. We would then have to calculate the density conditioning on the event actually occurring. 


There are two main methods used in survival analysis for modelling these durations:

\begin{enumerate}
\item With a Cox Proportional hazards model, in which the model covariates affect the duration through the hazard function $h(t)$
\item With an Accelerated Failure Time (AFT) model, in which the model covariates affect the duration by shifting the baseline distribution of $\tau$
\end{enumerate}

For a thorough description of both see \cite{kalbfleisch2011statistical}. In this paper we use an AFT model, which has the distinctive feature that the model covariates have a multiplicative influence on the survival time. In the simplest case, we can model the log of the i-th exceedance above the threshold level, using the following simple linear regression:

\begin{center}
$\log(\tau_i)=\mathbf{x}_i'\boldsymbol{\beta}+\varepsilon$
\end{center}

Then we have:

\begin{center}
$\tau_i=\exp{(\mathbf{x}_i'\boldsymbol{\beta})}\tau_{0,i}$
\end{center}

In this case, both the covariates and the parameters are fixed. Depending on the distribution we assume for $\varepsilon_t$, we get different parametric models. In this paper, we assume for simplicity that $\varepsilon \sim N(0,\sigma^2)$, and thus have a log-normal distribution for the durations. We considered other choices for the distribution of the duration random variables also, including the generalised gamma distribution. However, we chose the log-normal case as it is easier to fit and allows us to explore a large number of model structures.


In the log-normal case, the observation random variables have the following distribution function and survival function:
\begin{equation*}
f(t|\mathbf{x}_i)=\frac{1}{t\sqrt{2\pi\sigma^2}} \exp{[-\frac{(\log(t)-\mathbf{x}'_i\boldsymbol{\beta})^2} {2\sigma^2}]}
\end{equation*}
\begin{equation*}
S(t|\mathbf{x}_i)=1-F(t|\mathbf{x}_i)=\frac{1}{2}-
\frac{1}{2}\erf(\frac{\log(t)-\mathbf{x}'_i\boldsymbol{\beta}}{\sqrt{2\sigma^2}})
\end{equation*}

Define $u=\frac{\log(t)-\mathbf{x}'_i\boldsymbol{\beta}}{\sqrt{2\sigma^2}}$. Then the log likelihood is:
\begin{center}
$l(\boldsymbol{\beta},\sigma)=\log L(\beta, \sigma)=$

$\Sigma_{i \in U} \log(f(\tau_i)) + \Sigma_{k \in C}\log(S(\tau_k))=$

$\Sigma_{i \in U} [-\log(\tau_i \sqrt{(2\pi(\sigma)^2})-u_i^2] + \Sigma_{k \in C}\log(\frac{1}{2}-
\frac{1}{2}\erf(u_k))$
\end{center}

The partial derivatives with respect to $\beta_j$ and $\sigma$ are:
\begin{equation*}
\frac{\partial l}{\partial \beta_j}=
\Sigma_{i \in U} [2u_i\cdot \frac{x_{i,j}}{\sqrt{2(\sigma)^2}}]+
\Sigma_{k \in C} [\frac{\frac{1}{\sqrt{\pi}}\exp(-u_k^2)\frac{x_{k,j}}{\sqrt{2(\sigma)^2}}}{\frac{1}{2}-\frac{1}{2}\erf(u_k)}]
\end{equation*}

\begin{equation*}
\frac{\partial l}{\partial \sigma}=\Sigma_{i \in U} [\frac{-1+2u_i^2}{\sigma}]+
\Sigma_{k \in C} [-\frac{\frac{1}{\sqrt{\pi}}\exp(-u_k^2)(\frac{u_k}{\sigma})}
{\frac{1}{2}-\frac{1}{2}\erf(u_k)}]
\end{equation*}

We estimate the parameters via MLE, with a Newton gradient descent method, using standard optimisation packages in R.

The AFT regression framework assumes that we can relate the dependence of the durations on a combination of common covariates and threshold specific covariates directly, through a rescaling of time.  The sign of the coefficient for a given covariate indicates the direction of the partial effect of this variable on the conditional probability that the spread deviation duration will exceed a time $t$.

\subsection{Model LOB Covariates}
\label{covariates}
We consider the following covariates in our model structures. In the following, a `level' of the LOB is defined as one in which there is at least 1 resting limit order. Thus the first 5 levels of the bid are the 5 levels closest to the quote mid-point, where there is available volume for trading.  

\begin{itemize}
\item{The total number of asks in the first 5 levels of the LOB at time $t$, obtained according to $x_t^{(1)}=\sum_{i=1}^{5}\left |V_{t}^{a,i}\right |$} (where $\left | \cdot  \right |$ is the number of orders at a particular level), and is denoted $ask$ hereafter

\item{The total number of bids in the first 5 levels of the LOB at time $t$, obtained according to $x_t^{(2)}=\sum_{i=1}^{5}\left |V_{t}^{b,i}\right |$}, denoted $bid$

\item{The total ask volume in the first 5 levels of the LOB at time $t$, obtained according to $x_t^{(3)}=\sum_{i=1}^{5}TV_{t}^{a,i}$}, denoted $askVolume$

\item{The total bid volume in the first 5 levels of the LOB at time $t$, obtained according to $x_t^{(4)}=\sum_{i=1}^{5}TV_{t}^{b,i}$}, denoted $bidVolume$

\item{The number of bids $x_t^{(5)}$ in the LOB that had received price or size revisions (and were thus cancelled and resubmitted with the same order ID), denoted by $bidModified$}. 

\item{The number of asks $x_t^{(6)}$ in the LOB that had received price or size revisions, denoted by $askModified$}. 

\item{The average age (in ms) $x_t^{(7)}$ of bids in the first 5 levels at time $t$, denoted by $bidAge$.}
		
\item{The average age $x_t^{(8)}$ of asks in the first 5 levels at time $t$, denoted by $askAge$.}

\item The instantaneous value of the spread at the point at which the $i$-th exceedance occurs, which is given by $x_t^{(9)}=p_{t}^{a,1}-p_{t}^{b,1}-1$.

\item{For the 9 previously defined covariates, we also include exponentially weighted lagged versions. For example, in the case of the $x_t^{(s)}$ covariate, the respective lagged covariate value is then given by:
\begin{equation}
EWLx_t^{(s)}=\sum_{n=1}^d w^n x^{(s)}_{t - n\Delta}
\end{equation}
where for a time $t$, we consider $w=0.75$ is the weighting factor, $d=5$ is the number of lagged values we consider and $\Delta=1s$ is the interval between the lagged values. These covariates are hereafter denoted with the `l' prefix.}

\item The number of previous deviations in the interval $[t-\delta,t]$ above the threshold level, with $\delta=1s$, denoted by $prevexceed$. 

\end{itemize}

\section{Results and Discussion}
\label{sec:res}
\begin{figure}[!ht]
\begin{center}
\includegraphics[width=0.48\textwidth]{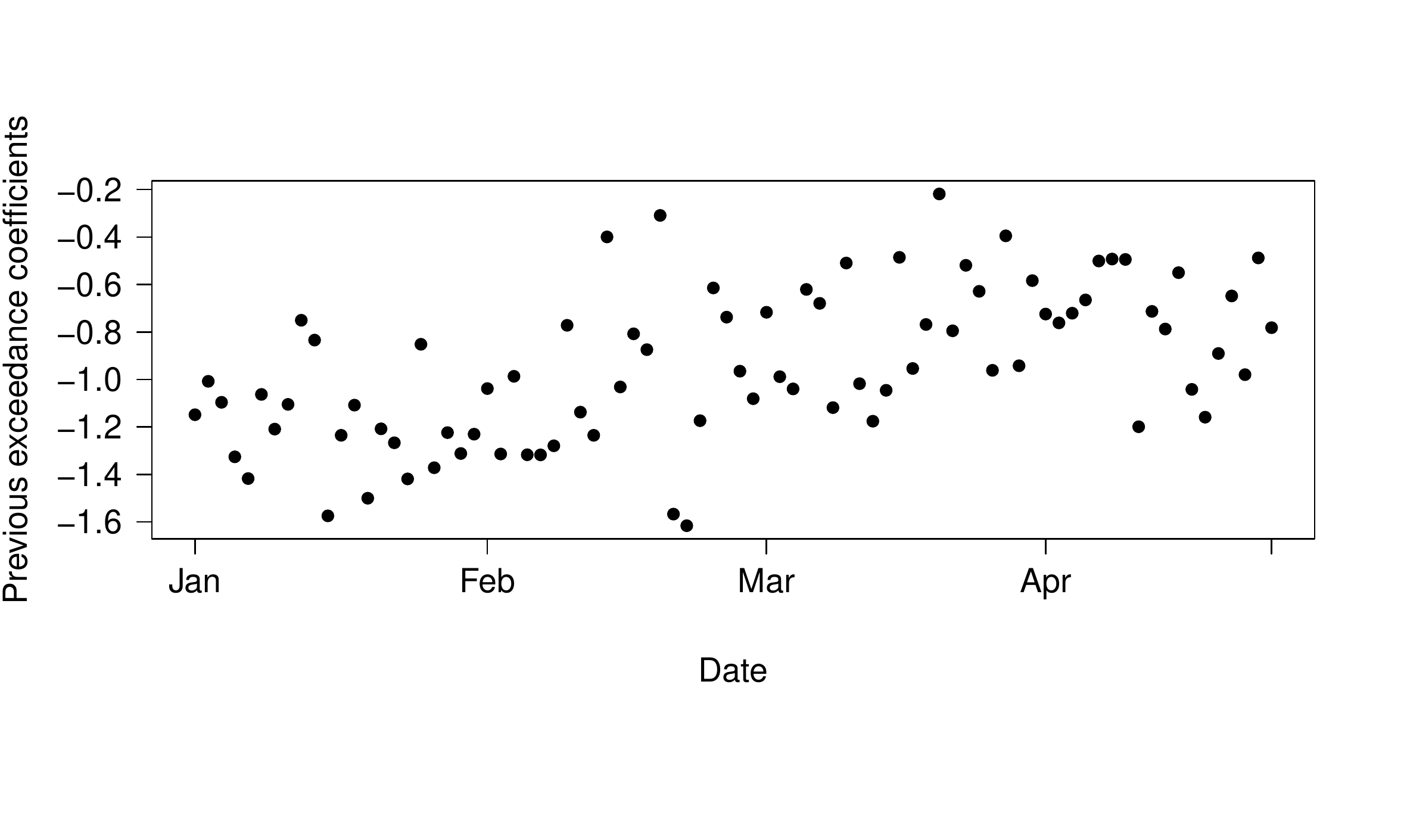}
\vspace{-3em}
\includegraphics[width=0.48\textwidth]{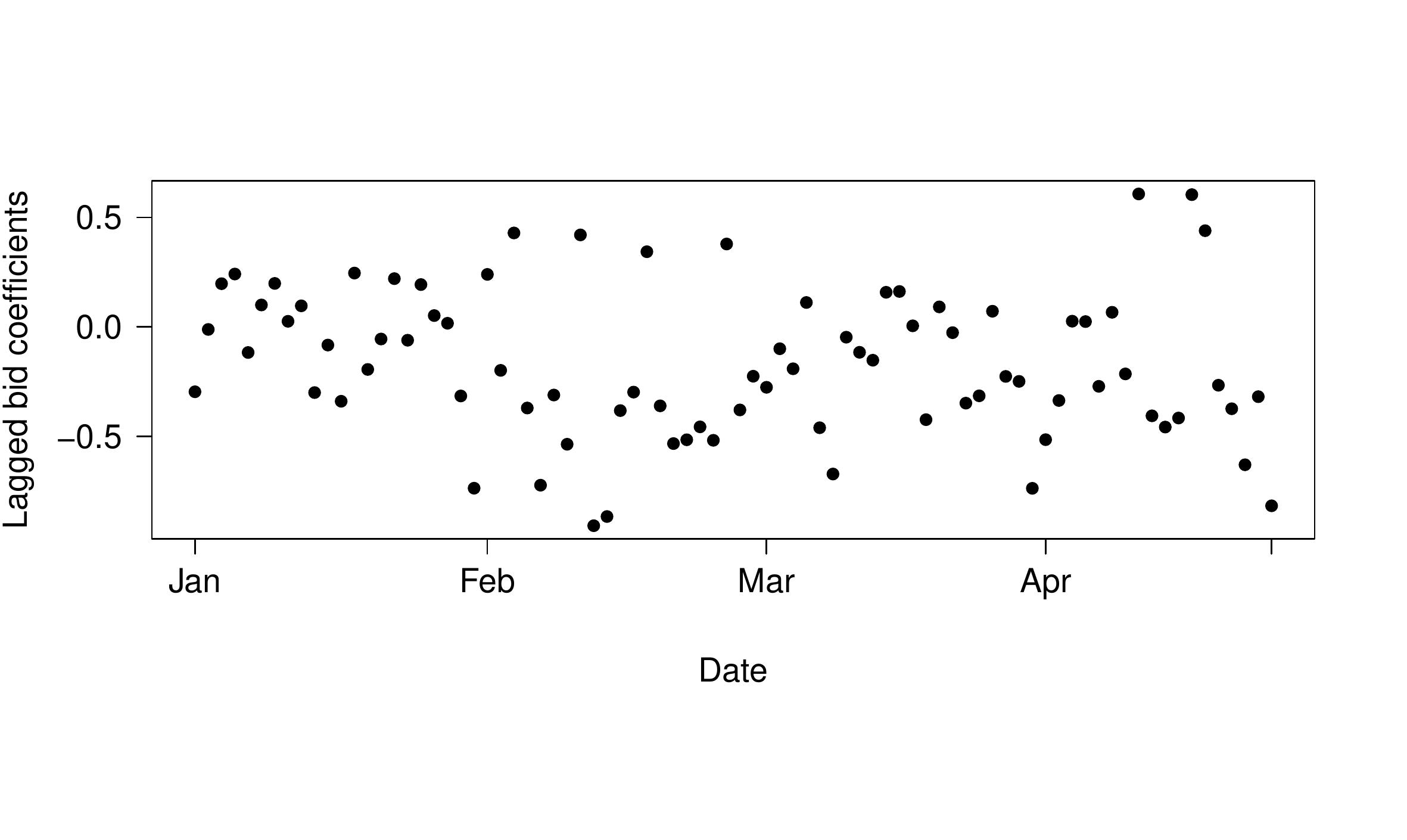}
\caption{The variation in the fitted daily model of the coefficients of the number of previous deviations and the (lagged) number of bids over our sample period }
\label{fig:coefs2}
\end{center}
\end{figure}

For the empirical evaluation of the explanatory power of our model regarding the variation in the durations of spread deviations, we adopt the AFT model formulation described in Section \ref{subsec:surv}. Our dataset consists of an 82-day trading sample, and, in order to fit our model, we can either assume stationarity of the spread deviation durations over the entire period, or only for a single day. We find that the former would not be a good assumption, as  the coefficient values for the fitted model vary over the period. For two of the covariates used to explain the duration of spread deviations, we show the variability of the coefficients over the four month period in Figure \ref{fig:coefs2}. The coefficients in the fitted daily models vary, and thus we cannot assume stationarity over the sample period. 


%
%

\subsection{Model selection}

In statistical modelling, one of the most prominent issues is finding the best regression equation, which entails choosing a subset of covariates that optimises some selection criterion \cite{gatu2006branch}. Including additional covariates always increases the explanatory power of a model, but may result in overfitting. A common approach used for model selection is thus to penalise the least squares of log likelihood scores, such that they take into account model size. This favours more parsimonious models and examples of criteria are Mallows' $C_p$, and Akaike's Information Criterion.  

\begin{figure}[ht!]
  \begin{center}
  \includegraphics[width=0.48\textwidth]{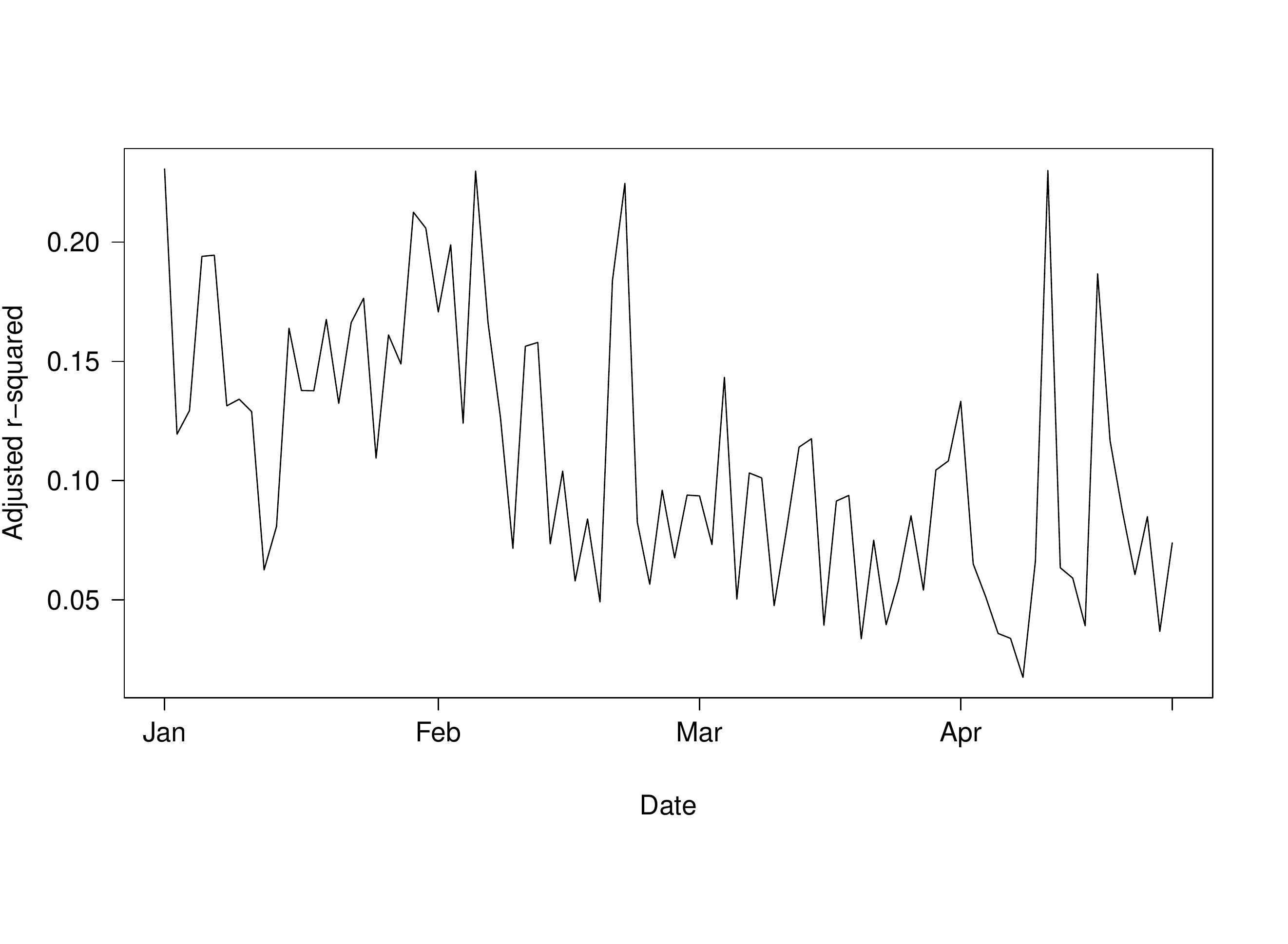}
  \vspace{-4em}
	\caption{The adjusted r-squared values over time}
	\label{adjr2}
	\end{center}
\end{figure}

The explanatory performance of our models was assessed in terms of the adjusted coefficient of determination (adjusted $R^2$), which we briefly explain here. The coefficient of determination is
\begin{equation*}
R^2=\frac{SS_e}{SS_t}
\end{equation*} 
which corresponds to the total variation explained by the regression model, where $SS_e$ and $SS_t$ are, respectively, the explained sum of squares and the total sum of squares. When introducing additional explanatory variables, we would always expect the $R^2$ value to increase. The adjusted $R^2$ is often used in its place, as it penalises larger models:
\begin{equation*}
R^2_{adj}=1-(1-R^2)\frac{N-1}{N-k-1}
\end{equation*} 

We find that our model has substantial explanatory power, when compared to regression models aiming to explain the variation in LOB quantities of interest. We show in Figure \ref{adjr2} that the adjusted $R^2$ results obtained from fitting the model every day for the four month period are above 15\% on many days, with scores above 20\% on some days also. As this is the result for the full model fit, we should be able to improve on this result also, by selecting the subset that maximises the adjusted $R^2$. 

This, however, poses a computational problem. In a regression model with $p$ covariates that can be included in a model, we have $2^p-1$ possible models to choose from. As $p$ increases, an exhaustive search of the entire space of possible models would thus be exponential in $p$. Although strategies to improve the efficiency of this search have been discussed, e.g. in \cite{gatu2006branch}, for a large value of $p$, an exhaustive search through all possible models is prohibitive in terms of computational power.
 
In order to search through the model space, we thus employ a modification of the leaps package in R \cite{lumley2004leaps}, which uses an efficient version of the branch-and-bound algorithm first described in \cite{furnival1974regressions}. 
The algorithm can offer vast performance improvements, by eliminating large sections of the search space. It is guaranteed to terminate, yielding the subset that maximises our selection criterion. 

A brief description of the general algorithm is as follows: For a given set of models in a partitioned model space, the algorithm proceeds by calculating upper and lower bounds for the selection criterion, for a supermodel and submodel of that set, respectively. If, during the search process, another model has been identified that has a higher selection criterion score than the upper bound, the given set can then safely be ignored, as it cannot give rise to a better performing model. Otherwise, the set is partitioned further. This process and partitioning is repeated until we have a singleton model, which is then evaluated.    

\begin{figure}[ht!]
  \begin{center}
	\includegraphics[width=0.48\textwidth]{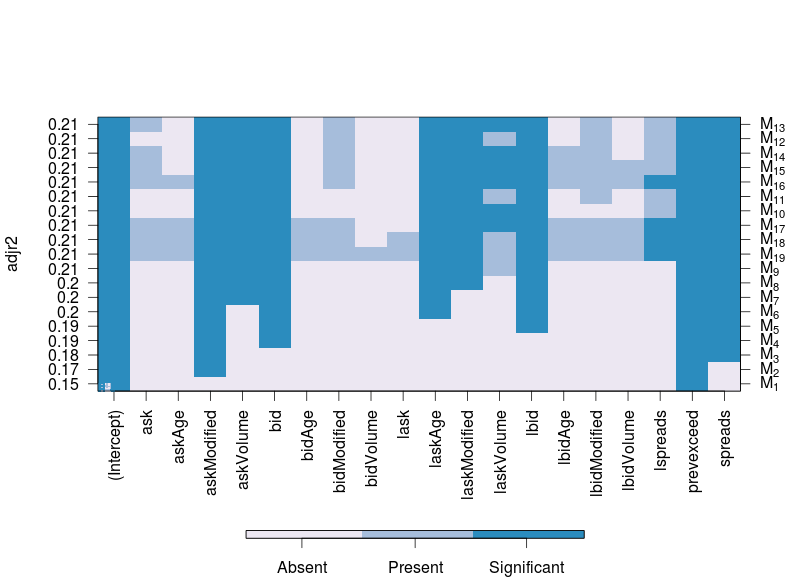}
	\caption{The best model for each subspace $M_1 \ldots M_p$ fit to a single trading day for stock Alstom SA. The models are ranked by the best adjusted-$R^2$ value, and we see that in this case, the best scoring model is obtained using a subset of 13 covariates. We differentiate between covariates that were found to be significant, or not. Out of the 13 covariates in the best model, only 9 are found to be significant at the 5\% level. }
	\label{subsetplot}
	\end{center}
\end{figure}

In our case, for every model subspace $M_i,i=1 \ldots p$, where $M_i$ contains all possible models with $i$ parameters - $C\binom{p}{i}=\frac{p!}{i!(p-i)!}$ models in total - we are searching for the model that maximises the adjusted $R^2$ criterion. Our modification to the leaps package is in the presentation of the results, so that it distinguishes between covariates that are selected to be part of the model (`Present'), and covariates that are significant in particular models (`Significant'), in Figure \ref{subsetplot}. 

From this, we observe the covariates that are consistently present as we move across subspaces. This is interesting because it gives us a relative measure of the contribution of that covariate across different assumptions of parsimony for the model. Particularly for higher model subspaces, some of the covariates in each subset model are not significant, and we distinguish between the covariates that are significant or not, at a 5\% level of significance.

The best models for each model subspace are ranked by the adjusted $R^2$ value, although there are very small differences in the best model in the first 11 rows (they only differ at most in the third decimal point of the score). The vertical lines in the graph represent the covariates that are consistently part of the best model for every subspace. We observe that the spread and number of previous deviations are covariates that are consistently part of the best model for every subspace.

\begin{figure}[ht!]
	\begin{center}
	\includegraphics[width=0.48\textwidth]{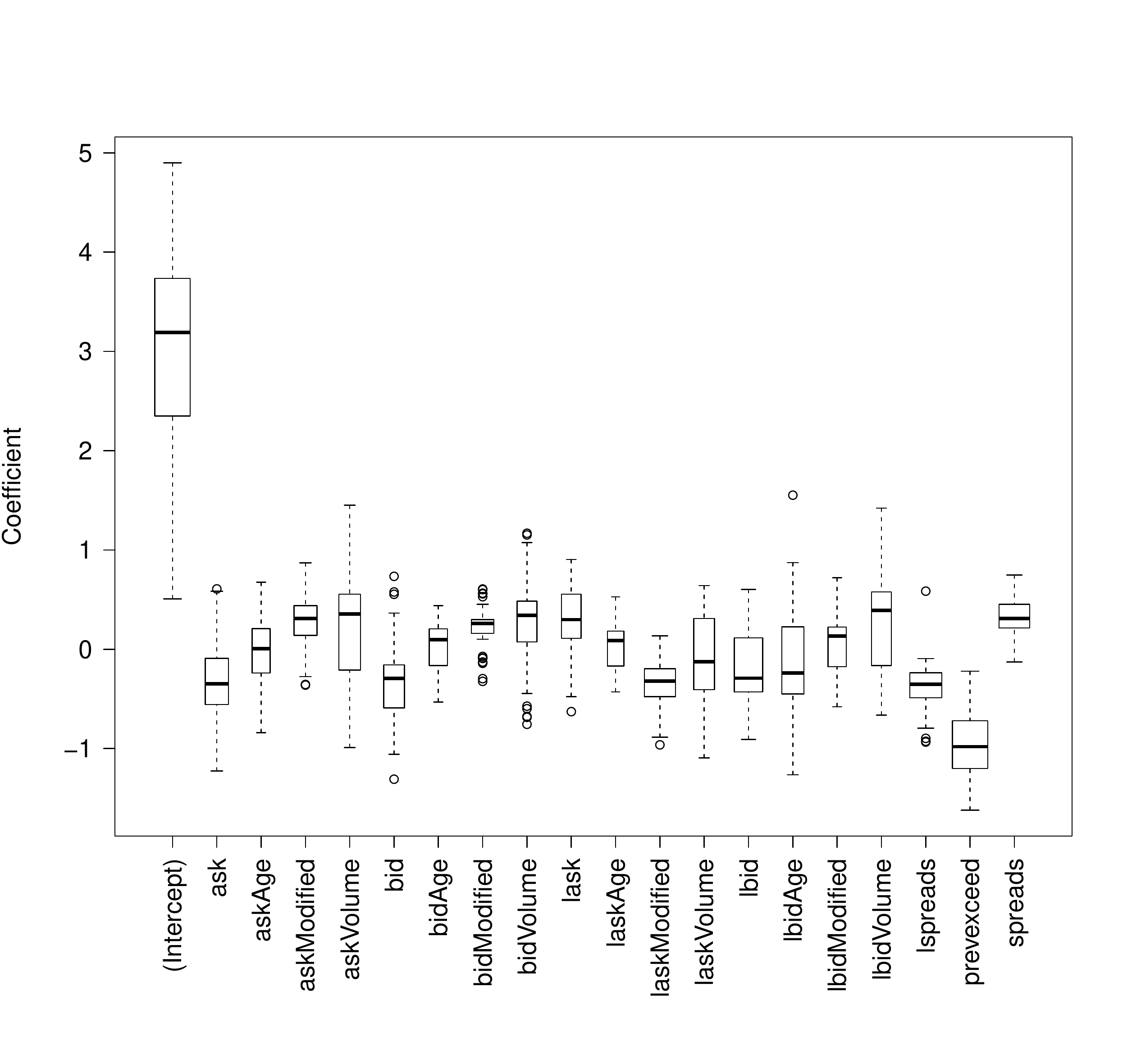}
	\vspace{-3em}
	\caption{Coefficients of the best models of any size (in terms of the adjusted $R^2$ values) for Alstom SA. The width of every boxplot is proportional to the square root of the number of times that the covariate appears in the best model over the four month period.}
	\label{fig:bestcoefsboxplot}
	\end{center}
\end{figure}

In order to compare the relative effect of each covariate, we first normalise the covariate values so that they have the same mean and standard deviation. We perform the analysis above for every day in our dataset, and select the best performing (in terms of the adjusted $R^2$ score) model each time. We show in Figure \ref{fig:bestcoefsboxplot} the values of each coefficient in the fitted models over time, also indicating the frequency with which each covariate is selected to be part of the best performing model.

Regarding the values of the coefficients, the coefficient of the `spreads' covariate, which tracks the spread immediately after a deviation even, is generally positive, indicating that it is associated with an increase in the duration of spread deviations. On the other hand, the coefficient of the number of previous observations (spread deviations) in the last second is negative, indicating that it is associated with a decrease in the duration of spread deviations. Finally, we note that the coefficients of covariates tracking the number of orders in the first 5 levels that have had price or size revisions (`bidModified' and `askModified') are also found to be generally positive. 

These results match our intuition: It would take a larger spread a longer time to return to a threshold value. In addition, we would expect the duration of the deviation to be shorter, when the spread has been fluctuating around that level in the near past. We see in Figure \ref{fig:bestcoefsboxplot} that `spreads' and `prevexceed' covariates are most frequently found to be significant in explaining the variation in the observation variables.

\begin{figure}[ht!]
	\begin{center}
	\includegraphics[width=0.48\textwidth]{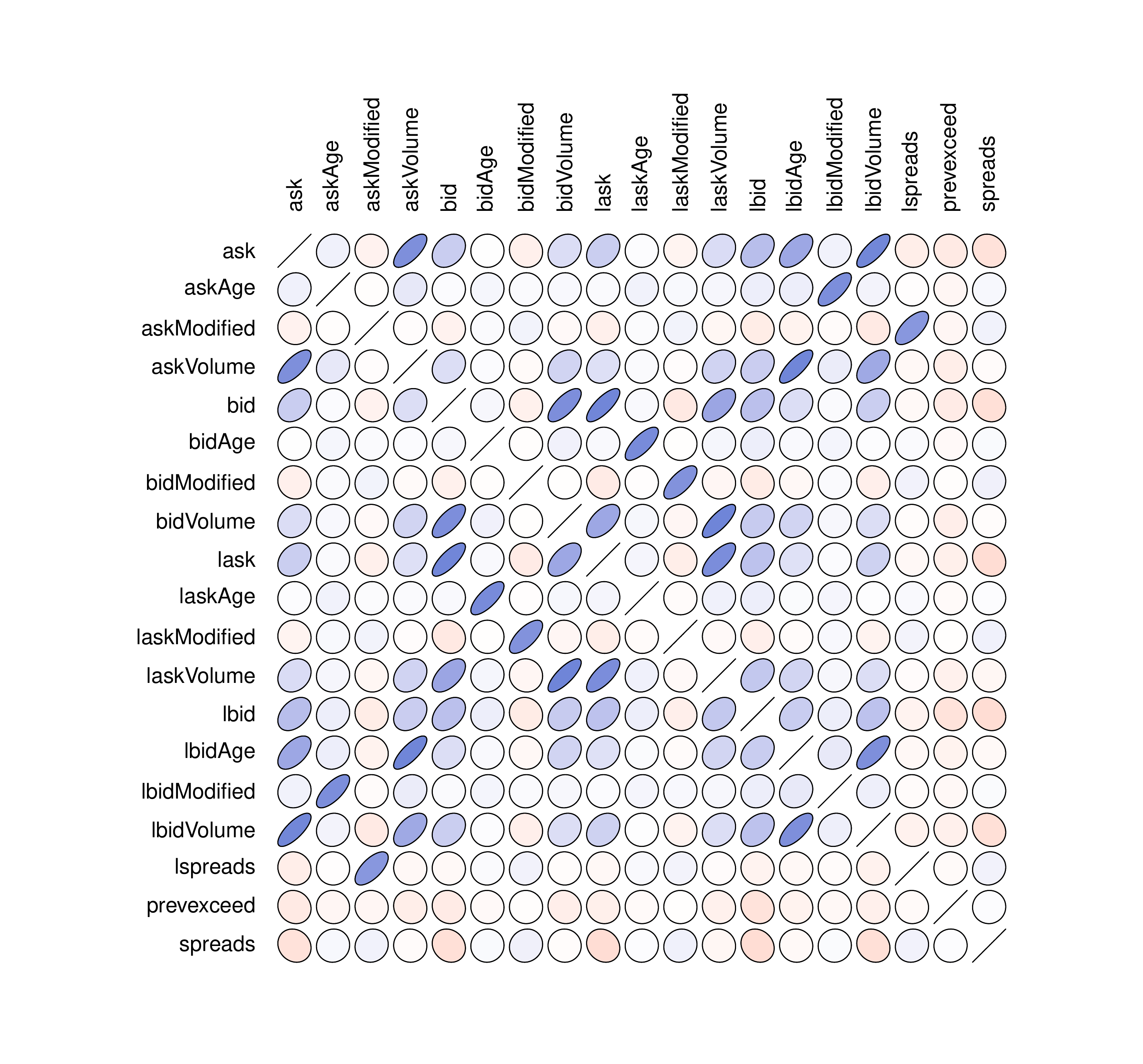}
	\vspace{-3em}
	\caption{Correlation plot of the covariates in the model for the entire 77-day dataset for stock Vallourec SA. The blue ellipses indicate positive correlations, while red ellipses indicate negative ones, and the narrower the ellipse, the stronger the correlations between the two covariates. }
	\label{corplot}	
	\end{center}
\end{figure}

We observe a reduction in the number of covariates found to be consistently significant in higher model subspaces. A possible explanation for the covariates being less significant in these subspaces would be that some are positively correlated, leading to multicollinearity and a reduction in the significance of individual covariates. To assess this, we calculate all pairwise correlations between the covariate  for the four month period , which we present in Figure \ref{corplot}, and we indeed find that there are some pairs that are positively correlated.

\section{Conclusion}
\label{sec:conc}
We have presented a flexible survival regression framework to model the duaration of intra-day spread fluctuations. This allows one to incorporate variables from the LOB and to make short-term predictions about spread deviations under different LOB regimes. Importantly, we have shown that such a class of models has good explanatory power in capturing features of these durations. In addition, we show that the conditional relationship between the spread deviation durations and the LOB structure is non-stationary inter-daily, as reflected by the change in the coefficients of the explanatory covariates. In the future, one may wish to introduce additional features to this modelling framework, in order to capture the time-series structure.

The covariates found to contribute most to explaining the variation in these durations were the size of the spread when a deviation event occurred and the number of previously observed deviations in the last second. The former was found to increase the expected duration of the deviation, an indication that the LOB takes longer to recover from a larger shock to the spread. The number of orders in the LOB that have had price or size revisions were also found to contribute positively to the duration. In contrast, the number of previous deviations in the recent past was found to be associated with a swifter return to the threshold level.

These covariates, amongst others, were consistently selected in the estimation of the best fitting models in a range of model subspaces and were statistically significant at the 5\% level. Thus, these covariates could, for example, form the basis of recommendation for models of trade execution, where the trading algorithm could estimate the time between subsequent tranches of a large order.



\bibliographystyle{IEEEtranS}

\bibliography{all.bib,Liquidity_Bibliography.bib}
\end{document}